\documentclass{elsart}
\usepackage{natbib}
\usepackage{graphics}
\usepackage{graphicx}
\usepackage{amssymb}
\def\accepted@prefix{; accepted~} 

\begin{document}
\begin{frontmatter}

\title{Orbital Phase Spectroscopy of four High Mass X-ray Binary Pulsars to 
Study the Stellar Wind of the Companion}
\author{Sachindra Naik$^{a}$, U. Mukherjee$^{b}$, B. Paul$^{c}$ \& C. S. Choi$^{d}$}

\address[label1]{Astronomy \& Astrophysics Division,
		 Physical Research Laboratory, Navrangpura,
		 Ahmedabad - 380 009, India}
		
\address[label2]{B P Poddar Institute of Management and Technology
137,V.I.P. Road, Poddar Vihar, Kolkata-700 052, India}

\address[label3]{Astronomy \& Astrophysics Group,
		Raman Research Institute,
		Sadashivnagar, C. V. Raman Avenue,
		Bangalore--560 080, India}

\address[label4]{Center for Astrophysics, 
		Korea Astronomy and Space Science Institute, 
		61-1, Hwaam-dong, Yuseong-gu, Daejeon, 
		Republic of Korea--305 348}

\ead{snaik@prl.res.in, uddipan.mukherjee@gmail.com, bpaul@rri.res.in, cschoi@kasi.re.kr}

\begin{abstract}
Our work focuses on a comprehensive orbital phase dependent spectroscopy
of the four High Mass X-ray Binary Pulsars (HMXBPs) 4U~1538-52, GX~301-2, 
OAO~1657-415 \& Vela~X-1. We hereby report the measurements of the variation 
of the absorption column density and iron-line flux along with other spectral 
parameters over the binary orbit for the above-mentioned HMXBPs in elliptical 
orbits, as observed with the Rossi X-ray Timing Explorer (RXTE) and the 
BeppoSAX satellites. A spherically symmetric wind profile was used as a 
model to compare the observed column density variations. Out of the four 
pulsars, only in 4U 1538-52, we find the model having a reasonable 
corroboration with the observations, whereas in the remaining three the 
stellar wind seems to be clumpy and a smooth symmetric stellar wind model 
appears to be quite inadequate in explaining the data. Moreover, in GX 301-2, 
neither the presence of a disk nor a gas stream from the companion was 
validated. Furthermore, the spectral results obtained in the case of 
OAO~1657-415 \& Vela~X-1 were more or less similar to that of GX~301-2.

\end{abstract}


\end{frontmatter}

\section{The Pulsars, Observations \& Analysis}
{\bf 4U~1538--52} was first detected with the UHURU satellite
(Giacconi et al. 1974). The spin period and orbital period of the pulsar 
were first estimated to be 529 s and 3.73 days, respectively, from the 
observations with Ariel 5 and OSO-8 (Davision et al. 1977). Using data 
from the RXTE, the eccentricity of the binary orbit was calculated to 
be $\sim$ 0.18 (Mukherjee et al. 2006). The X-ray spectrum has a 
prominent iron K-line (Makishima et al. 1987) and a pulse phase-dependent 
cyclotron resonance feature at 20 keV (Clark et al. 1990). It has a 
moderate X-ray luminosity of $\sim$ 10$^{36}$ erg s$^{-1}$. The optical 
counterpart was found to be an early B type supergiant star (QV~Nor) 
with H$\alpha$ emission lines (Parkes et al. 1978). The mass-loss rate 
and the terminal wind velocity were estimated as $\sim$10$^{-6}$ M$_{\odot}$ 
yr$^{-1}$ and $\sim$1000 km s$^{-1}$ respectively. 

{\bf GX 301--2 (4U 1223-62)} was discovered by White et al. (1976).
Using data from BATSE observations, the orbital period and eccentricity of 
the binary system were determined as $\sim$41.5 days and $\sim$0.46, 
respectively (Koh et al. 1997). The companion star Wray~977 has a 
B1 Ia+ spectral classification with a mass of 39 M$_{\odot}$ (Kaper et al. 
2006). The mass-loss rate and terminal velocity of the stellar wind are 
10$^{-5}$ M$_{\odot}$ yr$^{-1}$ and 305 km s$^{-1}$, respectively.
GX~301-2 shows a variable X-ray luminosity in the range (2-400) $\times$
$10^{35}$ erg s$^{-1}$, depending on the amount of the stellar wind captured,
which in turn depends on the density and velocity of the wind.

{\bf OAO~1657--415} was discovered by the Copernicus satellite
(Polidan et al. 1978). White \& Pravdo (1979) detected a
$\sim$38~s pulsation period for the neutron star. A steady spin-up
 time scape of 125 yr with short-term fluctuations are observed
from the period measurements from RXTE and BATSE (Baykal 2000).
BATSE observations also aided the discovery of X-ray eclipses by the 
stellar companion and the determination of a $\sim$ 10.44 day orbital 
period (Chakrabarty et al. 1993). They in turn used the orbital 
parameters to infer that the companion is a supergiant of spectral 
class B0--B6. OAO~1657-415 is unique among the known HMXBs in that 
it appears to occupy a transition region between mass transfer via 
a stellar wind and Roche lobe overflow (Chakrabarty et al. 1993).

{\bf Vela X-1} is also an eclipsing HMXBP with an orbital
period of $\sim$8.96 days (Barziv et al. 2001). It exhibits X-ray pulsations 
with a pulse period of 283 s (McClintock et al. 1976). The companion is an 
early-type primary star HD 77581, which is a massive B0.5Ib-–type supergiant 
(Brucato \& Kristian 1972). The inferred mass-loss rate of the stellar wind
of the companion is of the order of 10$^{-7}$ M$_\odot$ yr$^{-1}$ (Sako et 
al. 1999). Moreover, the terminal wind velocity has been determined to be 
1700 km s$^{-1}$ (Dupree et al. 1980). The pulsar orbits about the center 
of mass of the system at a distance of only about 0.6 stellar radii from 
the surface of the supergiant, which has a radius of 53.4 R$_\odot$.
This implies that the neutron star is deeply embedded within the influence
of the stellar wind. The typical X-ray luminosity of Vela X-1 is 4 $\times$ 
10$^{36}$ erg s$^{-1}$, but large flux variations have been observed 
(Kreykenbohm et al. 1999).

We observed 4U~1538-52 with RXTE from 2003-07-31 to 2003-08-07
covering out of eclipse phases for two binary orbits. We also
used the archival data from BeppoSAX obtained between 1998-07-29
to 1998-08-01, covering one binary orbit. GX~301-2 was observed by 
RXTE first from 1996-05-10 to 1996-06-15 and secondly from 2000-10-12 
to 2000-11-19. For OAO~1657-415, we analyzed the archival data as 
observed with RXTE from 1997-10-31 to 1997-11-11. In addition to these, 
we also analyzed an archival BeppoSAX Medium Energy Concentrator 
Spectrometer (MECS) observation taken on 2001-08-14 for $\sim$104 ks 
exposure. Vela~X-1 was observed with the Proportional Counter Array 
(PCA) of RXTE from 2005-01-01 to 2005-01-09. For the RXTE data, we 
took Standard-2 data products of the PCA and extracted the source and 
background spectra using the tool saextrct v 4.2d. The BeppoSAX data 
products were extracted from the MECS and Low Energy Concentrator 
Spectrometers (LECS) using circular regions of radius 4$'$ and 8$'$ 
respectively. The background subtracted source spectra were analyzed 
with the spectral analysis package XSPEC v 11.2.0. Since PCU0 had 
lost the propane layer, we did not use the data from it in any further 
analysis for Vela~X-1. This was not a major issue for such a highly 
luminous pulsar since it did not affect the statistics.

For 4U~1538-52, the spectral model used was a simple power law along with 
a line-of-sight absorption, a high energy cut off and a Gaussian line with 
center energy $\sim$6.4 keV. However in case of GX~301-2, the  Partial 
Covering Absorber Model (PCAM, Endo et al. 2002) with a high energy 
exponential cutoff and two Gaussian functions for iron K$_\alpha$ and 
K$_\beta$ lines was found to fit the RXTE data well compared to other 
models. The PCAM is described as two different power law components with 
the same photon index but different normalizations, being absorbed by 
different column densities (N$_{H1}$ \& N$_{H2}$) respectively. The 
analytical form of the PCAM that we have used for spectral fitting is :
$N(E) = {e^{-\sigma(E)N_{\mathrm H1}}}(S_{1}+S_{2}e^{-\sigma(E)N_{\mathrm H2}}) {E^{-\Gamma}}$, 
where $N(E)$ is the intensity, $\Gamma$ is the photon index, $N_{\mathrm H1}$ 
and $N_{\mathrm H2}$ are the two equivalent hydrogen column densities, 
$\sigma$ is the photo-electric cross-section, $S_{1}$ and $S_{2}$ are the 
respective normalizations of the power law. For OAO~1657-415 \& Vela~X-1 
too, we observed that the PCAM was providing the best fit for both MECS and 
the RXTE spectra. Moreover, since RXTE is a non--imaging instrument and 
OAO~1657-415 lies under the influence of the Galactic ridge emission, we 
had to explicitly incorporate the ridge emission as a separate background 
spectral component. We put that in XSPEC as a sum of Raymond-Smith plasma 
and a power-law with appropriate normalizations (Valinia \& Marshall 1998).

\begin{figure}
\vskip 9.5 cm
\includegraphics{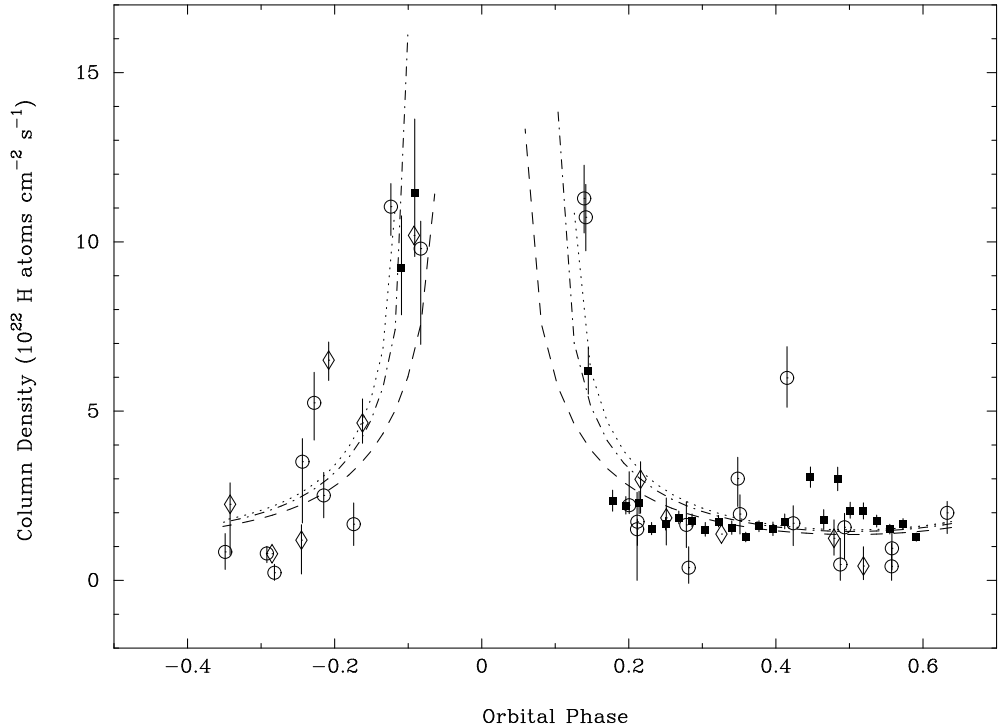}
\caption{Variation of Column Density versus Orbital Phase. The dashed 
line represents the model for inclination angle of 65$^\circ$, the 
dashed-dotted line for 75$^\circ$ and the dotted line for 85$^\circ$. 
Diamonds, filled squares, and circles denote measurements from the 
observations with RXTE in 1997, BeppoSAX in 1998, and RXTE in 2003 
respectively.}
\end{figure}

\section{Results \& Discussion}

{\bf 4U~1538-52} : The photon index, iron-line flux, cut-off energy and 
the e-folding energy measured with the pulse average spectrum taken over 
2-3 ks did not show any substantial variation along the orbit which 
suggests that the continuum X-ray spectrum of the pulsar is hardly 
affected during its revolution. An important detection was a notably 
variation in N$_{H}$. It shows a smooth variation over orbital phase, 
increasing gradually by an order of magnitude as the pulsar approaches 
eclipse (Fig 1 : mid-eclipse is defined by phase zero). At orbital phases 
far from the eclipse, the column density has a value of $\sim$ 1.5 $\times$ 
10$^{22}$ atoms cm$^{-2}$. Moreover, we compare the observed column density 
profile with a model estimated by assuming a spherically symmetric Castor, 
Abbott \& Klein (CAK 1975) wind from the companion star. The velocity 
profile of the line-driven wind is: 
$v_{\mathrm wind} =  {v_{\infty}}\sqrt{1-\frac {{R_\mathrm c}}{r}}$, 
where $v_{\infty}$ is the terminal velocity for the stellar wind, 
R$_{\mathrm c}$ is the radius of the companion and $r$ is the radial 
distance from center of the companion star. The column density profiles 
were derived using a numerical integration along the line of sight from 
the pulsar to the observer. With a mass-loss rate of $\sim$10$^{-6}$ 
M$_\odot$ yr$^{-1}$ and $v_\infty$ $\sim$1000 km s$^{-1}$, the model 
calculations of N$_H$ for different inclination angles when superposed 
on the observed values, it shows fairly reasonable agreement for three 
different inclination angles 65$^\circ$, 75$^\circ$ and 85$^\circ$ 
respectively (Fig. 1), indicating that a CAK wind from the companion star 
may produce the observed orbital dependence of the column density for a 
certain range of the orbital inclination ($>$65$^\circ$). We note here
that, in Fig. 1, we have not done any fitting of the measured column 
densities and the model calculated column densities at different orbital 
phases are rather plotted together. The phase resolved column density values
for the wind density model used here depend on the rate of mass loss, the
terminal velocity and the inclination angle of the binary orbit. The
data presented here, however, is not suitable for such a detailed analysis
but only shows the consistency of the model with the observations.

\begin{figure}
\vskip 10.9 cm
\includegraphics{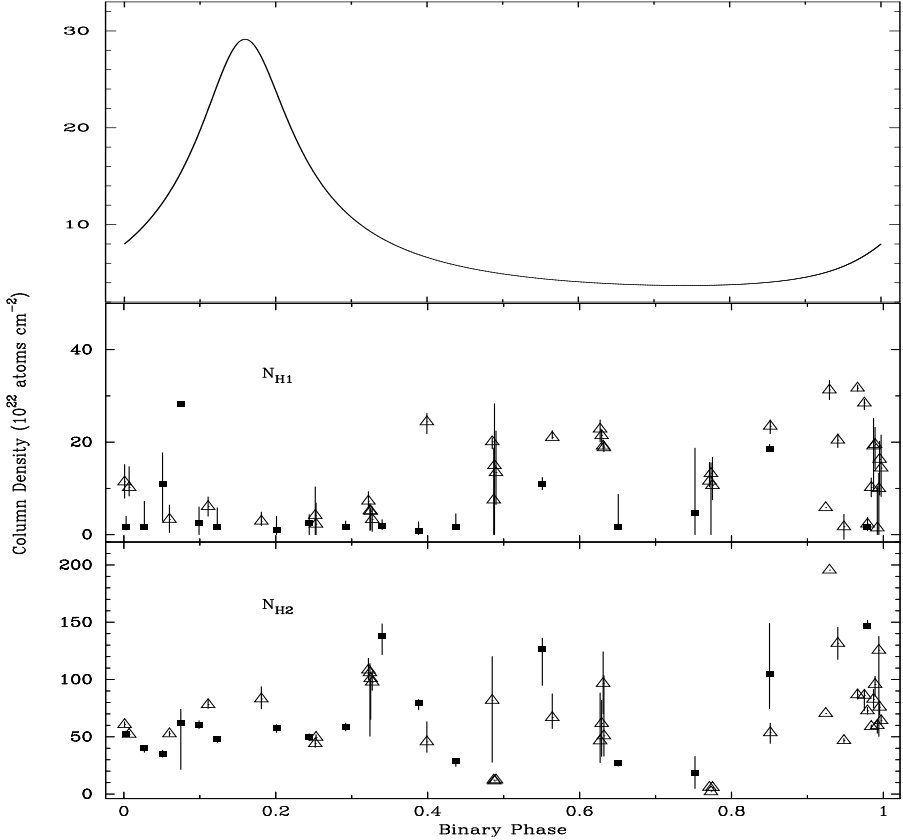}
\caption{The figure shows the N$_H$
variation for GX~301-2 with the uppermost panel depicts the model variation,
the middle panel showing the observed variation of N$_{\mathrm H1}$ and the
lower panel shows the observed variation of N$_{\mathrm H2}$. The error-bars
shown correspond to 90$\%$ confidence interval.}
\end{figure}

{\bf GX~301-2} : 
As in 4U~1538-52, the continuum parameters do not show any orbital modulation 
in GX~301-2. In this case, the variation of N$_{H1}$ \& N$_{H2}$ with orbital 
phase was not smooth. The values were very high with a large variation 
throughout the binary orbit (from 10$^{22}$ to 10$^{24}$ atoms cm$^{-2}$), 
indicating a clumpy nature of the stellar wind (Fig. 2). It is also seen 
that the covering fraction (defined as  Norm2/[Norm1+Norm2]) 
remains substantially high almost throughout the orbit which means that 
there is dense and clumpy material present throughout. This is also
supported by the detection of a strong Compton recoil component detected 
with the Chandra grating spectrum and its successful reproduction by Monte 
Carlo simulations (Watanabe et al. 2003). Thus it is clear that the 
observed variation in column density cannot be explained by a spherically 
symmetric CAK wind only, indicating the presence of strong inhomogeneities 
in the wind. There are two proposed models in this regard : a gas-stream 
model by Leahy (1991) and the equatorial disk model by Pravdo \& Ghosh 
(2001). However, the orbital dependence of the absorption column 
densities measured by us is very different from their predictions. Now, 
for the the iron K-line fluxes, we obtained peaks near periastron (Fig. 3).
The line equivalent width had a correlation with the column density 
(N$_{H2}$), suggesting that most of the iron line is produced by the 
local clumpy matter surrounding the neutron star.

\begin{figure}
\begin{center}
\includegraphics[angle=-90,width=12 cm]{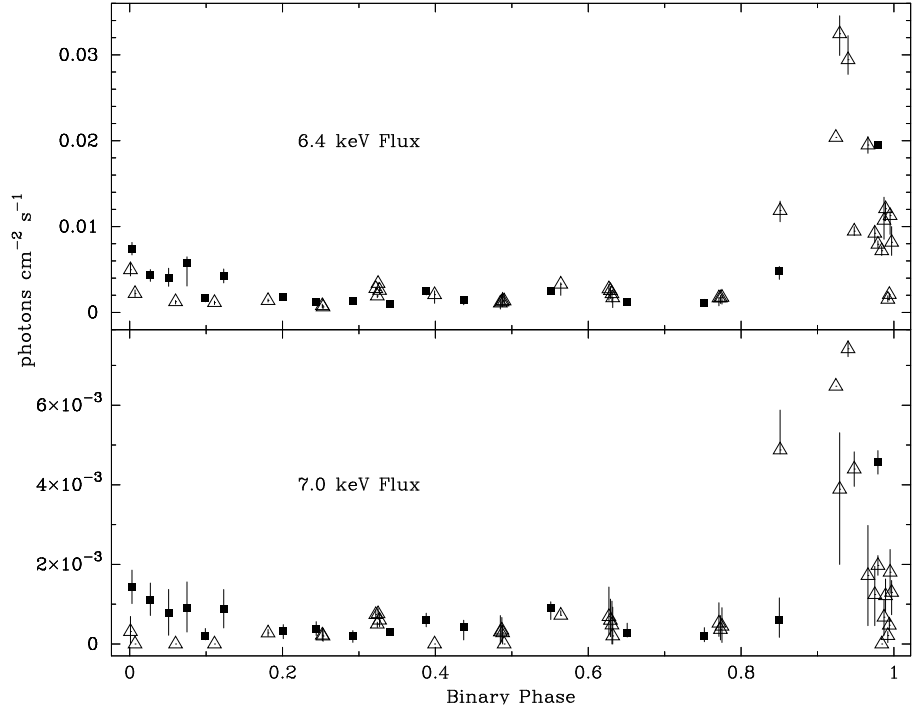}
\caption{The figure shows the variation of the iron-line flux with the 
orbital phase in GX~301-2. The upper and lower panels depict the flux 
corresponding to the 6.4 ${\mathrm keV}$ and 7.0 ${\mathrm keV}$ iron-lines 
respectively.}
\end{center}
\end{figure}

{\bf OAO~1657-415 \& Vela~X-1} : The average values of the free spectral 
parameters measured here over the out-of-eclipse phases of the 
binary orbit; viz. photon index, e-folding energy and cut-off energy,
do not exhibit any modulation due to the revolution of both the pulsars. 
These results seem to be in conformity with that obtained for GX~301-2 
\& 4U~1538-52. The column density (N$_{H2}$) of the material that absorbs 
the primary X-ray emission is found to be high for both the pulsars (Fig 4). 
A large variation of the column densities throughout the out-of-eclipse 
phases (from 10$^{22}$ to 10$^{24}$ atoms cm$^{-2}$) characterizes their 
X-ray spectra. There were instances of moderate to high values of covering 
fraction. Here, it should be mentioned that the number of out-of-eclipse 
measurements for OAO~1657-415 are probably not sufficient to arrive at a 
strong conclusion.

We do not attempt to compare the variations of the observed column densities
with the CAK model variation as we had done in case of 4U~1538-52 and GX~301-2 
since it is clear from the measurements that such a comparison would definitely
rule out the said model. In this context, we may point out that the spectral 
measurements for these two pulsars are similar in case of GX~301-2 (Mukherjee 
\& Paul 2004). For GX~301-2 though, formation of clumped blobs of matter could 
be accounted for the unusually low wind velocity and high mass loss rate, but
for the present case, such an inference is probably not possible. Moreover, 
unlike GX~301-2, we do not observe any systematic orbital modulation for the 
Iron-line fluxes. Furthermore, the iron-line equivalent width also does 
not show any definite correlation with N$_{H2}$.

\begin{figure}
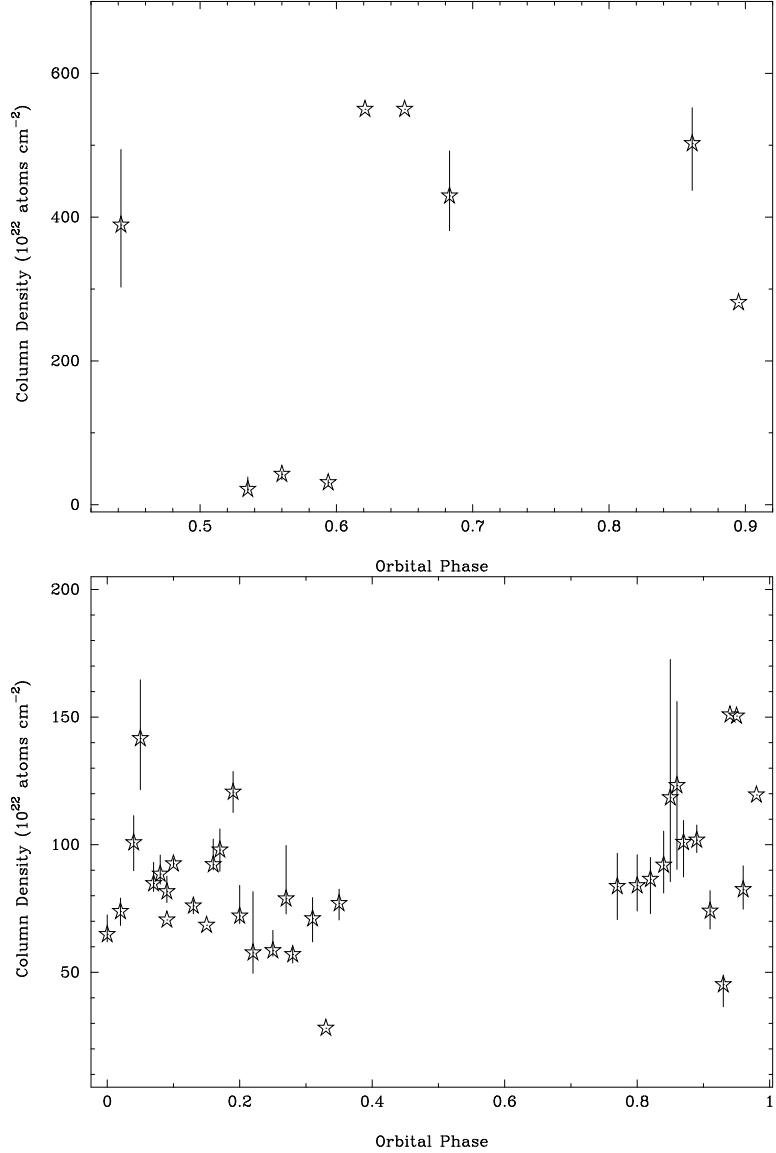

\centering
\includegraphics[width=3in,height=4in,clip,angle=-90.0]{fig4a.ps}
\includegraphics[width=3in,height=4in,clip,angle=-90.0]{fig4b.ps}
\caption{The figure shows the variation of N$_{H2}$ with orbital orbital
phase for OAO~1657-415 (top) and Vela~X-1 (bottom). It may be observed that
sufficient number of observations are lacking for OAO~1657-415. The error-bars
shown correspond to 90$\%$ confidence interval.}
\end{figure}

In view of the above exposition, we may conclude as follows :

\begin{enumerate}

\item The PCAM appears to be somewhat generic for pulsars which have variable
column density, at least when observed with RXTE or BeppoSAX.

\item For highly luminous supergiant binaries, the CAK Model of stellar wind 
does not suffice to describe the column density variations. This is 
corroborated by the results of GX~301-2, OAO~1657-415 \& Vela~X-1 in this 
work and of 4U~1700--37 (Haberl et al. 1989) and 4U~1907+09 (Roberts et al. 
2001).

\item Moderately luminous pulsars probably validate the spherical wind model, 
as in the case of 4U~1538-52 (this work) and X~1908+075 (Levine et al. 2004).
\end{enumerate}

\end{document}